\documentclass[onecolumn,prd,aps,12pt]{revtex4}
\usepackage{amsmath}
\usepackage{amsfonts}
\usepackage{amssymb}
\usepackage{graphicx}

\begin{document}

\title{An investigation of the $K_{F}$-type Lorentz-Symmetry Breaking Gauge
Models with Vortex-like Configurations. }
\author{H. Belich $^{a}$, F.J.L. Leal$^{b}$, H.L.C. Louzada$^{a}$, M.T.D.
Orlando$^{a}$}
\affiliation{$^{a}${\small {Universidade Federal do Esp\'{\i}rito Santo (UFES),
Departamento de F\'{\i}sica, Av. Fernando Ferrari 514, Vit\'{o}ria, ES, CEP
29060-900, Brasil;}}}
\affiliation{{\small {~}}$^{b}${\small Instituto Federal de Educa\c{c}\~{a}o, Ci\^{e}ncia
e Tecnologia do Estado do Esp\'{\i}rito Santo (IFES) - Campus Linhares, Av.
Filog\^{o}nio Peixoto S/N, Bairro Aviso, Linhares - ES, CEP 29901-291,
Brasil.}}
\email{belichjr@gmail.com, nandojll@ig.com.br, haofisica@bol.com.br,
mtdorlando@gmail.com }
\date{\today}

\begin{abstract}
For the CPT-even case of the minimal Standard Model Extension, the
spin-projector method is adopted to account for the breaking (tensor) $%
K_{\mu \nu \kappa \lambda }$ term. We adopt a particular decomposition of
this term in fourvectors, and carry out a detailed analysis of causality and
unitarity. From this study, we are able to impose conditions on the
decomposition of the $K_{\mu \nu \kappa \lambda }$ and vortex formation is
also investigated in different situations.
\end{abstract}

\pacs{11.30.Cp, 12.60.-i, 11.27.+d, 11.10.Lm.}
\maketitle

\section{Introduction}

\ Studies about symmetry breaking are well-known in nonrelativistic quantum
systems involving phase transitions such as ferromagnetic systems, where the
rotation symmetry is broken when the system is under the influence of a
magnetic field. Similarly, in superconductors the spontaneous violation of a
gauge symmetry shields electromagnetic interaction, but in type II
superconductors the magnetic field penetrates, as determined by Abrikosov 
\cite{ANO}, and form a 2D vortices lattice. The two-dimensional electronic
system is one of the most studied systems. Especially, a great amount of
effort has been done to investigate a 2D electron system under a strong
magnetic field to understand the quantum Hall effect. The Chern-Simons term
for the gauge field yield a change in statistics of the vortex \cite{Ezawa,
nagaosa} in these systems. Vortex configurations in planar models can be
induced by a Chern-Simons term. Such a type of solution presents a
interesting property to have electric charge \cite{CS}, \cite{CSV}. The
Chern-Simons vortices were studied with nonminimal coupling \cite{CSV1}, and
remain being a topic of intensive investigation with recent developments 
\cite{CSV2,Bolog}.

For relativistic systems, the study of symmetry breaking can be extended by
considering a background set up by tensors with rank $n\geq 1$. The\
background fields, in this situation, break the symmetry $SO\left(
1,3\right) $\ instead of the symmetry $SO\left( 3\right) $. This line of
research is known in the literature as the spontaneous violation of the
Lorentz symmetry \cite{extra3,extra1,extra2}. This new possibility of
spontaneous violation was first suggested in 1989 in a work of Kostelecky
and Samuel \cite{extra3} indicating that, in the string field theory, the
spontaneous violation of symmetry by a scalar field could be extended to
tensor fields. This extension has as an immediate consequence: a spontaneous
breaking of the Lorentz symmetry. In the electroweak theory, an $SU(2)$%
-doublet scalar field acquires a nonzero vacuum expectation value which
yields mass to the $SU(2)$ gauge bosons (Higgs Mechanism). Similarly, in a
string scenario a tensor field may trigger symmetry breaking. Nowadays,
these theories are encompassed in the framework of the Extended Standard
Model (SME) \cite{col} as a possible extension of the minimal Standard Model
of the fundamental interactions. For instance, the violation of the Lorentz
symmetry is implemented in the fermion section of the Extended Standard
Model by two CPT-odd terms: $v_{\mu }\overline{\psi }\gamma ^{\mu }\psi $\
and $b_{\mu }\overline{\psi }\gamma _{5}\gamma ^{\mu }\psi $, where $v_{\mu }
$\ and $b_{\mu }$\ correspond to the Lorentz-violating backgrounds.

\ An extension to the Chern-Simons form is implemented by $v_{\mu }$\
backgound field in the form, 
\begin{equation}
\Sigma _{CS}=-\frac{1}{4}\int dx^{4}\epsilon ^{\mu \nu \alpha \beta }v_{\mu
}A_{\nu }F_{\alpha \beta }.  \label{0.1}
\end{equation}

Such a term appears in the gauge sector of the SME and corresponds to a
CPT-odd sector (the Carroll-Field-Jackiw term \cite{Jackiw}). A possibility
to find an effect of the violation of this symmetry would be by the analysis
of the defects that could be formed after Lorentz-symmetry breaking has been
realized. Taking this point of view, we examine vortex solution in this
scenario.

The vortex solution of the Maxwell-Chern-Simons with Lorentz violation was
first studied in \cite{baeta}, and a dimensional reduction was adopted to
study planar vortex solution. As expected, this vortex solution presents an
electric charge. Also interference effects in topological defects with
Lorentz violation have been investigated\cite{knut}. This line of
investigation contemplates the possible phases that can be generated, and by
means of interference processes, we could detect the violation spontaneous
of Lorentz symmetry. As the expectation of spontaneous Lorentz symmetry
breaking by a background is beyond of the dynamics of the Standard Model, by
experimental measurements we expect to set up stringent bounds on the
parameters of the breaking ( $v_{\mu }$, $b_{\mu }$).

Besides the CPT-odd terms, in the gauge sector, we have the CPT-even sector,
which is represented by a tensor $K^{\mu \nu \alpha \beta }$\ with the same
symmetries as the Riemann tensor, as well as an additional double-traceless
condition \cite{klin}. In this set-up, we present two possibilities of
constructing a supersymmetric version for the $K$-type models.

We actually propose to carry out the supersymmetric extension to the bosonic
action below:

\begin{equation}
S=-\frac{1}{4}\int d^{4}x\;K_{\mu \nu \kappa \lambda }F^{\mu \nu }F^{\kappa
\lambda }.  \label{acao1}
\end{equation}

\ The CPT-even gauge sector of the SME has been studied since 2002, after
the pioneering contributions by Kostelecky \& Mewes \cite{19, 20}, and there
is extensive literature dealing with the extension of the standard model in
the even sector of (SSM) by this term \cite{cas}. We propose to work with a
decomposition vectorial above because we can access to new physical
properties where CPT violation does not occur.

The \textquotedblleft tensor\textquotedblright\ $K_{\mu \nu \kappa \lambda }$%
\ is CPT even, i. e., it does not violate the CPT-symmetry. Though CPT
violation implies violation of Lorentz invariance \cite{greenberg}, the
reverse is not necessarily true. The action above is Lorentz-violanting in
the sense that the \textquotedblleft tensor\textquotedblright\ $K_{\mu \nu
\kappa \lambda }$\ has a non-zero vacuum expectation value. That
\textquotedblleft tensor\textquotedblright\ presents the following
symmetries:

\begin{equation}
K_{\mu \nu \kappa \lambda }=K_{\left[ \mu \nu \right] \left[ \kappa \lambda %
\right] },\;K_{\mu \nu \kappa \lambda }=K_{\kappa \lambda \mu \nu },\;K^{\mu
\nu }{}_{\mu \nu }=0=0,  \label{anz1}
\end{equation}

we can reduce the degrees of freedom and take into account the ans\"{a}tze 
\cite{klin}:

\begin{equation}
K_{\mu \nu \kappa \lambda }=\frac{1}{2}\left( \eta _{\mu \kappa }\tilde{%
\kappa}_{\nu \lambda }-\eta _{\mu \lambda }\tilde{\kappa}_{\nu \kappa }+\eta
_{\nu \lambda }\tilde{\kappa}_{\mu \kappa }-\eta _{\nu \kappa }\tilde{\kappa}%
_{\mu \lambda }\right) ,  \label{anz2}
\end{equation}

\begin{equation}
\tilde{\kappa}_{\mu \nu }=\kappa \left( \xi _{\mu }\xi _{\nu }-\eta _{\mu
\nu }\xi ^{\alpha }\xi _{\alpha }/4\right) ,  \label{anz3}
\end{equation}

\begin{equation}
\kappa =\frac{4}{3}\tilde{\kappa}^{\mu \nu }\xi _{\mu }\xi _{\nu },
\end{equation}

where $\tilde{\kappa}^{\mu \nu }$\ is a traceless \textquotedblleft
tensor\textquotedblright . Using the restrictions (\ref{anz2}), (\ref{anz3}%
), in expression (\ref{acao1}), we obtain,

\begin{equation}
S=\frac{\kappa }{4}\int d^{4}x\left\{ \frac{1}{2}\xi _{\mu }\xi _{\nu }F{%
^{\mu }}_{\kappa }F^{\kappa \nu }+\frac{1}{8}\xi _{\lambda }\xi ^{\lambda
}F_{\mu \nu }F^{\mu \nu }\right\} .  \label{acao2}
\end{equation}

An interesting topic of research is the discussion on the violation of
supersymmetry (susy) \cite{susy}. We have investigated the possibility that
SUSY and Lorentz symmetry are broken down at the same time. This study has
already been carried out for the odd sector, and we are proposing here the
extension to the even sector. This work is the beginning of a study of
consistency to this decomposition suggested in the sector even, and we are
also starting up an investigation of susy violation in this sector \cite{fer}%
.

In this work, we analyze the possibility of having a consistent quantization
of an Abelian theory which incorporates the Lorentz violating term of
equation (\ref{acao1}), whenever gauge spontaneous symmetry breaking (SSB)
takes place. Using the decomposition (\ref{anz2}, \ref{anz3}) the analysis
is carried out by pursuing the investigation of unitarity and causality as
read off from the gauge-field propagators. We therefore propose a discussion
at tree-approximation, without going through the canonical quantization
procedure for field operators. In this investigation, we concentrate on the
analysis of the residue matrices at each pole of the propagators. Basically,
we check the positivity of the eigenvalues of the residue matrix associated
to a given simple pole in order that unitarity is respected at a
semi-classical level.

In order to deepen our comprehension of the physics presented by this model,
we also study vortex-like configurations, by analyzing the influence of the
direction selected by $K_{\mu \nu \kappa \lambda }$\ in space-time. The
decomposition of the $K_{\mu \nu \kappa \lambda }$\ tensor produces
interesting modifications on the equations of motion that may yield vortex
formation.

This work is outlined as follows: in Section 2, we study the SSB and present
our method to derive the gauge-field propagators. In Section 3, we set our
discussion on the poles and residues of the propagators. We study the
formation of vortices in Section 4, and, finally, in Section 5, we present
our Concluding Comments.

\section{The Gauge-Higgs Model}

We propose to carry out our analysis by starting off from the action 
\begin{equation}
\Sigma =\int d^{4}x\left\{ -\frac{1}{4}F_{\mu \nu }F^{\mu \nu }+\left(
D_{\mu }\varphi \right) ^{\ast }D^{\mu }\varphi -V\left( \varphi \right) +%
\mathcal{L}_{\kappa }\right\} ,  \label{1}
\end{equation}%
where $L_{\kappa }$\ is the Lorentz violation term%
\begin{equation}
\mathcal{L}_{\kappa }=-\frac{1}{4}\left( K^{\mu \nu \rho \sigma }\,F_{\mu
\nu }F_{\rho \sigma }\right) .  \label{2}
\end{equation}%
\ Taking into account the ansatz presented in the introduction, we obtain :

\begin{equation}
K^{\mu \nu \rho \sigma }\,F_{\mu \nu }F_{\rho \sigma }=2\kappa \left( g^{\mu
\rho }\left( \xi ^{\nu }\xi ^{\sigma }-g^{\nu \sigma }\,\xi ^{e}\xi
_{e}/4\right) \right) F_{\mu \nu }F_{\rho \sigma }.
\end{equation}%
The potential, $V$, given by 
\begin{equation}
V(\varphi )=m^{2}\left\vert \varphi \right\vert ^{2}+\lambda \left\vert
\varphi \right\vert ^{4}
\end{equation}%
is the most general Higgs-like potential in 4D. Setting suitably the
parameters such that the $\varphi $-field acquires a non-vanishing vacuum
expectation value (v.e.v.), namely, $\lambda >0$\ and $m^{2}<0$, the mass
spectrum of the photon can be read off after the spontaneous breaking of
local gauge symmetry and the $\varphi $-field has been shifted by its v.e.v.
. The Higgs field is minimally coupled to the electromagnetic by means of
its covariant derivative under U(1)-local gauge symmetry, namely 
\begin{equation}
D_{\mu }\varphi =\partial _{\mu }\varphi +ieA_{\mu }\varphi .  \label{4}
\end{equation}

This symmetry is spontaneously broken, and the new vacuum is given by 
\begin{equation}
\langle 0|\varphi |0\rangle =a,
\end{equation}%
where 
\begin{equation}
a=\left( -\frac{m^{2}}{2\lambda }\right) ^{1/2};\,\,\,\,m^{2}<0.
\end{equation}%
As usually, we adopt the polar parametrization 
\begin{equation}
\varphi =\left( a+\frac{\sigma }{\sqrt{2}}\right) e^{i\rho /\sqrt{2}a},
\label{7}
\end{equation}%
where $\sigma $, and $\rho $\ are the scalar quantum fluctuations. Since we
are actually interested in the analysis of the excitation spectrum, we
choose to work in the unitary gauge, which is realized actually by setting $%
\rho =0$. Then, the bilinear gauge action is given as below: 
\begin{equation}
\Sigma _{g}=\int d^{4}x\left\{ -\frac{1}{4}F_{\mu \nu }F^{\mu \nu }-\frac{1}{%
4}2\kappa \left( \left( g^{\mu \rho }\left( \xi ^{\nu }\xi ^{\sigma }-g^{\nu
\sigma }\,\xi ^{e}\xi _{e}/4\right) \right) F_{\mu \nu }F_{\rho \sigma
}\right) +\frac{M^{2}}{2}A_{\mu }A^{\mu }\right\} ,
\end{equation}%
where $M^{2}=2e^{2}a^{2}$.

We can express this action as a bilinear as follows below:

\begin{equation}
\Sigma _{g}=\int d^{4}x\frac{1}{2}A^{\mu }\left\{ \mathcal{O}_{\mu \nu
}\right\} A^{\nu }
\end{equation}

where $O_{\mu \nu }$\ is the wave operator. The wave operator can be
formulated in terms of spin-projection operators as follows, where $\theta
_{\mu \nu }$\ and $\omega _{\mu \nu }$\ are respectively the transverse and
longitudinal projector operators: 
\begin{equation}
\theta _{\mu \nu }=g_{\mu \nu }-\frac{\partial _{\mu }\partial _{\nu }}{\Box 
}\;,\;\;\;\omega _{\mu \nu }=\frac{\partial _{\mu }\partial _{\nu }}{\Box }.
\end{equation}%
In order to invert the wave operator, one needs to add up other two new
operators, since the ones above do not form a closed algebra, as the
expression below indicates: 
\begin{equation}
\Sigma _{\mu \nu }=\xi _{\mu }\partial _{\nu }\;,\;\;\lambda \equiv \Sigma
_{\mu }^{\;\mu }=\xi _{\mu }\partial ^{\mu }\;,\;\;\Lambda _{\mu \nu }=\xi
_{\mu }\xi _{\nu }.
\end{equation}%
We can express $O_{\mu \nu }$ as,

\begin{equation*}
\mathcal{O}_{\mu \nu }=\left( \left( 1-\kappa \,\xi ^{e}\xi _{e}/2\right)
\Box +\kappa \lambda ^{2}+M^{2}\right) \theta _{\mu \nu }+\left( \kappa
\lambda ^{2}+M^{2}\right) \left( \omega _{\mu \nu }\right) +\kappa \Box \xi
_{\mu }\xi _{\nu }-\kappa \lambda \left( \xi _{\mu }\partial _{\nu
}+\partial _{\mu }\xi _{\nu }\right) .
\end{equation*}%
The propagator is given by 
\begin{equation}
\left\langle 0\right\vert T\left[ A_{\mu }\left( x\right) A_{\nu }\left(
y\right) \right] \left\vert 0\right\rangle =-i\left( \mathcal{O}^{-1}\right)
_{\mu \nu }\delta ^{4}\left( x-y\right) .  \label{11}
\end{equation}%
These results indicate that two new operators, namely, $\Sigma $\ and $%
\Lambda $, must be included in order to have an operator algebra with closed
multiplicative rule. The operator algebra is displayed in Table 1.

\begin{center}
\begin{tabular}{|c|c|c|c|c|c|}
\hline
& $\theta _{\,\,\,\,\,\nu }^{\alpha }$ & $\omega _{\,\,\,\,\,\nu }^{\alpha }$
& $\Lambda _{\,\,\,\,\,\nu }^{\alpha }$ & $\Sigma _{\,\,\,\,\,\nu }^{\alpha }
$ & $\Sigma _{\nu }^{\,\,\,\,\,\alpha }$ \\ \hline
$\theta _{\mu \alpha }$ & $\theta _{\mu \nu }$ & $0$ & $\Lambda _{\mu \nu }-%
\frac{\lambda }{\Box }\Sigma _{\nu \mu }$ & $\Sigma _{\mu \nu }-\lambda
\omega _{\mu \nu }$ & $0$ \\ \hline
$\omega _{\mu \alpha }$ & $0$ & $\omega _{\mu \nu }$ & $\frac{\lambda }{\Box 
}\Sigma _{\nu \mu }$ & $\lambda \omega _{\mu \nu }$ & $\Sigma _{\nu \mu }$
\\ \hline
$\Lambda _{\mu \alpha }$ & $\Lambda _{\mu \nu }-\frac{\lambda }{\Box }\Sigma
_{\mu \nu }$ & $\frac{\lambda }{\Box }\Sigma _{\mu \nu }$ & $\xi ^{2}\Lambda
_{\mu \nu }$ & $\xi ^{2}\Sigma _{\mu \nu }$ & $\lambda \Lambda _{\mu \nu }$
\\ \hline
$\Sigma _{\mu \alpha }$ & $0$ & $\Sigma _{\mu \nu }$ & $\lambda \Lambda
_{\mu \nu }$ & $\lambda \Sigma _{\mu \nu }$ & $\Lambda _{\mu \nu }\Box $ \\ 
\hline
$\Sigma _{\alpha \mu }$ & $\Sigma _{\nu \mu }-\lambda \omega _{\mu \nu }$ & $%
\lambda \omega _{\mu \nu }$ & $\xi ^{2}\Sigma _{\nu \mu }$ & $\xi ^{2}\Box
\omega _{\mu \nu }$ & $\lambda \Sigma _{\nu \mu }$ \\ \hline
\end{tabular}%
\ \vspace{2mm}

Table 1: Multiplicative table fulfilled by $\theta ,\omega ,\,S,\,\Lambda $\
and $\Sigma $. The products are supposed to obey the order
\textquotedblright\ column times row\textquotedblright .
\end{center}

Using the spin-projector algebra displayed in Table 1, the propagator may be
obtained after a number of algebraic manipulations. Its explicit form in
momentum space can be written down upon use of the equation:

\begin{equation*}
O_{\mu \alpha }\left( \mathcal{O}^{-1}\right) _{\nu }^{\alpha }=\theta _{\mu
\nu }+\omega _{\mu \nu }.
\end{equation*}

The expressions containing the poles of the propagator are cast below:

\begin{eqnarray}
D &=&\left( 1-\kappa \,\xi ^{2}/2\right) \Box +\kappa \lambda ^{2}+M^{2},
\label{pD} \\
E &=&\left( 1+\kappa \,\xi ^{2}/2\right) \Box +M^{2}.  \label{pE}
\end{eqnarray}

The final form of the propagator is

\begin{eqnarray}
\langle A_{\mu }A_{\nu }\rangle  &=&\frac{i}{D}\left\{ \theta _{\mu \nu
}+\left( \frac{1}{M^{2}}\left( \frac{\left( \left( 1-\kappa \,\xi
^{2}/2\right) \Box +M^{2}\right) ^{2}+\kappa \lambda ^{2}\left( 1+\kappa
\,\xi ^{2}/2\right) \Box }{E}\right) \right) \omega _{\mu \nu }\right.  
\notag \\
&&\left. -\left( \frac{\kappa \Box }{E}\right) \Lambda _{\mu \nu }+\left( 
\frac{\lambda \kappa }{E}\right) \Sigma _{\mu \nu }+\left( \frac{\lambda
\kappa }{E}\right) \Sigma _{\nu \mu }\right\} .  \label{13}
\end{eqnarray}

The expression above enables us to set up our discussion on the nature of
the excitations, which can be read off as pole propagators, present in the
spectrum. At a first sight, the denominator \ $E$\ appearing in connection
with the operators $\omega $, $\Lambda $, $\Sigma $, once multiplying the
overall denominator $D$, could be the origin for dangerous multiple poles
that plague the quantum spectrum with ghosts. For this reason, a careful
study of this question is worthwhile. With this purpose, it is advisable to
split our discussion into 2 cases: time-like, and space-like $\xi _{\mu }$.
In the next section, we\ carefully analyze these possibilities.

\section{Dispersion relations, stability and causality}

In this section, we analyze the causality from a classical perspective (the
tree-level), which is based on the positivity of the poles of the
propagators in the variable $p^{2}$,\ which is the associated momentum. The
starting point of our analysis is the propagator, whose poles are associated
with the dispersion relations (DR), which provide information on the
stability and causality of the model. The analysis of causality is related
to the signal of the poles propagator, given in terms of $p^{2},$\ so that
we should have $p^{2}\geq 0$\ to preserve causality (avoid tachyons). From
the viewpoint of second quantization, the stability is related to the states
of positive energy in the Fock space for any moment. Here, stability is
directly associated with positive energy for each mode that is coming from
to the DR. The propagators of the fields, given by expressions. ($D,$\ $E$\
) present two families of poles in $p^{2}$:

\begin{eqnarray}
(i)D &=&\left( 1-\kappa \,\xi ^{2}/2\right) \left( -p_{\mu }p^{\mu }\right)
+\kappa \lambda ^{2}+M^{2}, \\
(ii)E &=&\left( 1+\kappa \,\xi ^{2}/2\right) \left( -p_{\mu }p^{\mu }\right)
+M^{2}.
\end{eqnarray}

For $\xi ^{\mu }=(1;0,0,0)$,\ when we analyze the dispersion relations: $(i)$%
\ gives us poles$,p^{0}=\pm \sqrt{\frac{\left( 2-\kappa \right) \left\vert 
\vec{p}\right\vert ^{2}+2M^{2}}{2+\kappa }}=\pm m_{t}$, and to the $(ii)$\
we have $\ p^{0}=\pm \sqrt{\left\vert \vec{p}\right\vert ^{2}+\frac{M^{2}}{%
\left( 1+\kappa /2\right) }}$. We note that if we make the substitution to $%
\vec{p}\rightarrow -\vec{p}$, the dispersion relations keep the same. This
behavior implies that we do not have birefringence. To avoid tachyonic modes
we have $\kappa \epsilon (-2,2).$

The case $\xi ^{\mu }=(0;0,0,1)$, $(i)$\ gives us poles $p^{0}=\pm \sqrt{%
\frac{2M^{2}+\left\vert \vec{p}\right\vert ^{2}\left( 2-\kappa \right) }{%
2+\kappa }}=\pm m_{s},$\ and, for $(ii)$,\ we have $\ p^{0}=\pm \sqrt{%
\left\vert p^{3}\right\vert ^{2}+\frac{M^{2}}{\left( 1-\kappa \,/2\right) }}$%
.

From these (DR) we see that the condition $\ \kappa \epsilon (-2,2)$\ avoid
presence of tachyons. Then upon control of the value of $\kappa $,\ we are
able to preserve causality.\ As the (DR) above do not exhibit a linear
dependence on the component $p^{0}$,\ the theory is not birefringent.

\section{Analysis of unitarity}

For the analysis of the model at the classical level, we adopt the method of
saturating the propagator with external currents. The fact that our model
has two sectors (the scalar, and "gauge") implies that we saturate the
scalar propagator and "gauge" separately. Thus, we write the propagators
saturated as: 
\begin{equation*}
SP_{\langle A_{\mu }A_{\nu }\rangle }=J^{\ast \mu }{\langle A}_{\mu }{(k)A}%
_{\nu }{(k)\rangle \ J}^{\nu }{.}
\end{equation*}

The continuity equation, $\partial _{\mu }J^{\mu }=0,$\ in the space of
momenta takes the form of: $k_{\mu }J^{\mu }=0$. To infer on the physical
nature of the simple pole, have to calculate the eigenvalues{}of the matrix
of residues at each of the poles. This will be done in the sequel. We must,
without loss of generality, set the external vector as ($\xi ^{\mu }$)
time-like, and space-like. We shall carry out an analysis of the residues by
taking $p^{\mu }=(p^{0};0,0,p^{3})$ as the linear momentum. The current
conservation law also reduces to two the number of terms of the photon
propagator which contribute to the calculation of the saturated propagator:

\begin{equation}
B_{\mu \nu }(k)=\frac{-i}{D}\left\{ g_{\mu \nu }-\left( \frac{\kappa \Box }{E%
}\right) \Lambda _{\mu \nu }\right\} ,
\end{equation}%
\ 
\begin{equation}
SP_{\langle A_{\mu }A_{\nu }\rangle }=J_{\mu }^{\ast }(k)\biggl\{\frac{-i}{D}%
\left( g_{\mu \nu }-\left( \frac{\kappa \Box }{E}\right) \Lambda _{\mu \nu
}\right) \biggr\}J_{\nu }(k).
\end{equation}%
\ Writing this expression in the space of momenta, we obtain: 
\begin{equation}
SP_{\langle A_{\mu }A_{\nu }\rangle }=J^{\ast \mu }(k)\biggl\{iB_{\mu \nu }%
\biggr\}J^{\nu }(k).
\end{equation}

Our present task consists in checking the character of the poles presented
in different configurations of $\xi ^{\mu }$. We pursue our analysis of the
residues by taking $p^{\mu }=(p^{0},0,0,0,p^{3}).$To infer about the
physical nature of the simple poles, we have to calculate the eigenvalues of
the residue matrix for each of these poles. This is done in the sequel.

With $\xi ^{\mu }=(0;0,0,1)$\ space-like, and in the pole : $p_{0}=m_{s}$,
and taking into account the current conservation, we have to study the
residues matrix of the form,

\begin{equation}
res_{p^{0}=m_{s}}B_{ij}(k)=\frac{i}{m_{s}}\left( 
\begin{array}{cccc}
0 & 0 & 0 & 0 \\ 
0 & \left( 2+\kappa \right)  & 0 & 0 \\ 
0 & 0 & \left( 2+\kappa \right)  & 0 \\ 
0 & 0 & 0 & \left( 2+\kappa \right) +2\left( \frac{\kappa \Box }{E}\right) 
\end{array}%
\right) ,
\end{equation}

and we observe that the only ambiguity in the sign of the matrix terms is in
the $res_{p^{0}=m_{s}}B_{33}.$Then we have to study the dependence of the
sign with $\kappa $. We call $W=res_{p^{0}=m_{s}}B_{33}$\ and,

\begin{equation}
W=\left( 2+\kappa \right) +\frac{\kappa \Box }{E}=\left( 2+\kappa \right) +%
\frac{\kappa \left( p^{3}\right) ^{2}-M^{2}}{\left( 1-\kappa /2\right)
\left( p^{3}\right) ^{2}+M^{2}}.
\end{equation}

To make our analysis independent of momentum, we take the limit $%
M\rightarrow 0$. Essentially, we are going to the limit in which the
condensate begins to show up. The expression of $W$\ in this limit is,

\begin{equation}
W=\frac{4-\kappa ^{2}+4\kappa }{\left( 2-\kappa \right) },
\end{equation}

and the interval in which $W>0$\ is $\kappa \epsilon (2-2\sqrt{2},2)$\ or \ $%
\kappa \epsilon (2+2\sqrt{2},+\infty )$. To avoid tachyonic modes, we impose 
$\kappa \epsilon (-2,2).$\ Then we select only the interval $\kappa \epsilon
(2-2\sqrt{2},2).$

With $\xi ^{\mu }=(1;0,0,0)$\ time like, and in the pole : $%
p_{0}^{2}=m_{t}^{2}$, we have,%
\begin{equation}
W=\left( 2+\kappa \right) +\frac{\kappa \left( p^{3}\right) ^{2}-M^{2}}{%
M^{2}+\left( 1-\kappa /2\right) \left( p^{3}\right) ^{2}}.
\end{equation}

In the limit $M\rightarrow 0$, we have

\begin{equation}
W=\frac{4-\kappa ^{2}+2\kappa }{2-\kappa },  \notag
\end{equation}%
\ 

and the valid interval in which $W>0$\ is $\kappa \epsilon (1-\sqrt{5},2)$.
We have to make $(1-\sqrt{5},2)\cap (2-2\sqrt{2},2)=(2-2\sqrt{2},2)$. Then
the interval of validity, for a while, is $\kappa \epsilon (2-2\sqrt{2},2).$

This preliminary study done in this section establishes the domain of
validity of $\kappa ,$ avoiding ghosts and tachyons. In the next section, we
shall study whether this model can provide vortex solutions obeying the
restriction of the $\kappa \epsilon (2-2\sqrt{2},2).$

\section{A discussion on vortex-like configurations}

Once our discussion on the consistency of the quantum-mechanical properties
of the model has been settled down, we would like to address to an issue of
a classical orientation, namely, the reassessment of vortex-like
configurations in the presence of Lorentz-breaking term as the one we tackle
here.

In our case, with the $K_{\mu \nu \kappa \lambda }$\ term included, we get,
from the action (\ref{1}), the equations of motion 
\begin{equation}
D^{\mu }D_{\mu }\varphi =-m^{2}\varphi -2\lambda \varphi |\varphi |^{2},
\end{equation}%
and 
\begin{equation}
\kappa \xi ^{\nu }\xi ^{\sigma }\partial ^{\rho }F_{\rho \sigma }-\kappa
g^{\nu \rho }\xi ^{\mu }\xi ^{\sigma }\partial _{\mu }F_{\rho \sigma
}+ie\left( \varphi \partial ^{\nu }\varphi ^{\ast }-\varphi ^{\ast }\partial
^{\nu }\varphi \right) +2e^{2}A^{\nu }\varphi ^{\ast }\varphi =-\left(
1-\kappa \,\xi ^{2}/2\right) \left( \partial _{\mu }F^{\mu \nu }\right) ,
\label{euu}
\end{equation}%
so that we can explicitly derive the modified Maxwell equations, 
\begin{eqnarray}
-\left[ \left( 1-\kappa \,\xi ^{2}/2-\kappa \left( \xi ^{0}\right)
^{2}\right) \mathbf{\nabla \cdot }-\kappa \lambda \vec{\xi}\cdot \right] 
\mathbf{E} &=&\kappa \xi ^{0}\left( \partial ^{0}\vec{\xi}\cdot \mathbf{E}+%
\vec{\xi}\cdot \mathbf{\nabla }\times \mathbf{B}\right) +  \notag \\
\  &&+ie\left( \varphi \partial ^{0}\varphi ^{\ast }-\varphi ^{\ast
}\partial ^{0}\varphi \right) +2e^{2}\varphi ^{\ast }\varphi \mathbf{\Phi ,}
\label{divE}
\end{eqnarray}

\begin{equation}
\nabla \times E=-\frac{\partial \mathbf{B}}{\partial t},
\end{equation}

and

\begin{equation}
\mathbf{\nabla .B=}0,
\end{equation}%
\begin{eqnarray}
-\left( 1-\kappa \,\xi ^{2}/2\right) \left( -\partial _{0}\mathbf{E+\nabla }%
\times \mathbf{B}\right)  &=&\kappa \xi ^{0}\left( \vec{\xi}\mathbf{\nabla .}%
-\kappa \lambda \right) \mathbf{E}-\kappa \left( \vec{\xi}+\kappa \lambda
\right) \left( \vec{\xi}\cdot \mathbf{\nabla }\times \mathbf{B}\right)  
\notag \\
&&-ie\left( \varphi \mathbf{\nabla }\varphi ^{\ast }-\varphi ^{\ast }\mathbf{%
\nabla }\varphi \right) +2e^{2}\mathbf{A}\varphi ^{\ast }\varphi .
\label{amp}
\end{eqnarray}

We would like to handle the modified Maxwell equations above (eqs. (\ref%
{divE})-(\ref{amp})), before going on to analyze vortex configurations. We
need to understand the anisotropy generated by the kind of Lorentz violation
we are considering. For this purpose, we remove the charged scalar field and
see that the modified Maxwell equations presents the contribution of the
fourvector $\xi ^{\mu }$\ decomposition. The modified Gauss law is, in the
stationary regime,

\begin{equation}
\left[ \left( 1-\kappa \,\xi ^{2}/2-\kappa \left( \xi ^{0}\right)
^{2}\right) \mathbf{\nabla \cdot }-\kappa \lambda \vec{\xi}\cdot \right]
E=\kappa \xi ^{0}\left( \vec{\xi}\cdot \mathbf{\nabla }\times \mathbf{B}%
\right) +2e^{2}\varphi ^{\ast }\varphi \Phi .
\end{equation}

To search for the vortex-type solutions, we consider a scalar field in $2$%
-dimensional space

\begin{equation}
\varphi =\chi \left( r\right) e^{in\theta }.
\end{equation}

The asymptotic solution is proposed to be a circle $(S^{1})$\ 
\begin{equation}
\varphi =ae^{in\theta };~~~~~~(r\rightarrow \infty ).  \label{38}
\end{equation}

\ Asymptotically, the magnetic field is screened and we have

\begin{equation}
\left( 1-\kappa \,\xi ^{2}/2-\kappa \left( \xi ^{0}\right) ^{2}\right) \frac{%
d}{dr}\left( \frac{d}{dr}\mathbf{\Phi }\right) -\kappa \lambda \left\vert
\xi _{r}\right\vert \frac{d}{dr}\Phi -2e^{2}a^{2}\Phi =0,
\end{equation}

the differential equation do not present a independent term in relation of $%
\Phi $. Then this equation admit the trivial solution $\Phi =0$, and this
imply that the vortex solution is not charged.

To seek the vortex-type solutions, we assume that the gauge field takes over
the form 
\begin{equation}
\mathbf{A}=\frac{1}{e}\mathbf{\nabla }(n\theta );~~~~~~(r\rightarrow \infty
),  \label{39}
\end{equation}%
or, in term of its components: 
\begin{equation}
A_{r}\rightarrow 0,~~~A_{\theta }\rightarrow -\frac{n}{er}%
;~~~~~~(r\rightarrow \infty ).  \label{40}
\end{equation}

Studying the modified Amp\`{e}re-Maxwell equation (\ref{amp}) in the
stationary regime, and as our vortex solution does not present electrical
charge ($E=0$), we have,

\begin{eqnarray}
-\left( 1-\kappa \,\xi ^{2}/2\right) \left( \mathbf{\nabla }\times \mathbf{B}%
\right)  &=&\kappa \vec{\xi}\left( \left( \mathbf{\nabla .}\left( \vec{\xi}%
\times \mathbf{B}\right) \right) \right) -\kappa \left( \left( \vec{\xi}%
\times \left( \vec{\xi}.\mathbf{\nabla }\right) \mathbf{B}\right) \right) + 
\notag \\
&&+ie\left( \varphi \mathbf{\nabla }\varphi ^{\ast }-\varphi ^{\ast }\mathbf{%
\nabla }\varphi \right) +2e^{2}A\varphi ^{\ast }\varphi .  \label{amp1}
\end{eqnarray}

In the case. $\xi ^{\mu }=(1;0,0,0),$

\begin{equation}
\frac{d}{dr}\left[ \frac{1}{r}\frac{d}{dr}(rA)\right] -2e\frac{\chi ^{2}}{%
\left( 1-\kappa \,/2\right) }\left( \frac{n}{r}+e\mathbf{A}\right) =0,
\end{equation}

in the approximation $\lim\limits_{r\rightarrow \infty }\chi (r)=a$, we
obtain the solution,

\begin{equation}
A(r)=-\frac{n}{er}+\frac{c}{e}CK_{1}\left( \left\vert e\right\vert \frac{a}{%
\sqrt{1-\kappa \,/2}}r\right) .\text{ }
\end{equation}

In the asymptotic limit,

\begin{equation}
\lim\limits_{r\rightarrow \infty }A(r)=-\frac{n}{er}+\frac{c}{e}\left( \frac{%
\pi }{2\left\vert e\right\vert \frac{a}{\sqrt{1-\kappa \,/2}}r}\right) ^{%
\frac{1}{2}}\exp \left( -\left\vert e\right\vert \frac{a}{\sqrt{1-\kappa \,/2%
}}r\right)   \label{vorttime}
\end{equation}

we have the asymptotic solution behavior is governed by $\frac{a}{\sqrt{%
1-\kappa \,/2}}=a^{\prime }(\kappa ).$\ In the interval $\kappa \epsilon (2-2%
\sqrt{2},2),$\ $a^{\prime }$\ is positive, and. in the $\lim\limits_{\kappa
\rightarrow 2^{-}}a^{\prime }(\kappa )=+\infty .$ What happens in this
limit? As $\kappa $ tends to $2^{-}$ the value of $a^{\prime }(\kappa )$\
increases, i. e., the condensate is enhanced obliging the vortex to be more
confined until it disappears.

To the case. $\xi ^{\mu }=(0;0,0,1)$,%
\begin{equation}
\frac{d}{dr}\left[ \frac{1}{r}\frac{d}{dr}(rA)\right] -2e\frac{\chi ^{2}}{%
\left( 1+\kappa \,/2\right) }\left( \frac{n}{r}+e\mathbf{A}\right) =0.
\end{equation}

The solution is in the asymptotic limit,%
\begin{equation}
\lim\limits_{r\rightarrow \infty }A(r)=-\frac{n}{er}+\frac{c}{e}\left( \frac{%
\pi }{2\left\vert e\right\vert \frac{a}{\sqrt{1+\kappa \,/2}}r}\right) ^{%
\frac{1}{2}}\exp \left( -\left\vert e\right\vert \frac{a}{\sqrt{1+\kappa \,/2%
}}r\right)   \label{vortspace}
\end{equation}

\begin{equation}
\kappa \epsilon (2-2\sqrt{2},2)
\end{equation}

In the interval $\kappa \epsilon (2-2\sqrt{2},2)$, $a^{\prime }=\frac{2a}{%
\sqrt{2+\kappa \,}}$\ is positive. When\ $\kappa $\ starts from the value $%
\left( 2-2\sqrt{2}\right) $, and goes up towards $2$,\ we observe that the
value of $a^{\prime }$\ decrease. Then, the vortex penetration increases,
since the condensate looses intensity, but the vortex is not completely
suppressed by the condensate.

To study the stability of the vortex solution, we calculate the energy
associated with this setup and we get the expression:

\begin{equation}
E=\frac{1}{8}B^{2}(4-\kappa )+\frac{M^{2}}{2}\left( A(r)\right) ^{2}
\end{equation}

the domain which we are considering, taking into account the criteria of
consistency, is\ $\kappa \epsilon (2-2\sqrt{2},2)$. In such range the energy
is positive, then we have a stable solution in the stationary regime.

\section{Concluding Comments}

Our work primarily makes the study of the quantization consistency of an
Abelian model with violation of Lorentz symmetry by the $K_{\mu \nu \kappa
\lambda }$ "tensor" ( decomposed in fourvectors $\xi ^{\mu }$)
contemporarily with the spontaneous breaking of gauge symmetry. Then, in
this work we access the Standard Model Extension in the even sector. The
analysis carried out with the help of the propagators, derived thanks to an
algebra of extended spin operators, reveals that unitarity is preserved if $%
\kappa \epsilon (2-2\sqrt{2},2)$. In this regime we have the foton with
three degrees of freedom.

Can such a phase transition produce topological defects? To answer that, we
have taken into account the classical vortex-like configurations. The
analysis of this defect shows interesting aspects: the interval of $\kappa $
from the analysis of consistence gives a stable solution. Taking into
account the solution (\ref{vorttime}), as $\kappa $ tends to $2^{-}$ the
value of $a^{\prime }(\kappa )$\ increases, i. e., the condensate is
enhanced by having the vortex be more, and more confined until it fades
off.\ On the other hand, taking the solution (\ref{vortspace}), as $\kappa $
tends to $2^{-}$ the value of $a^{\prime }(\kappa )$\ increases, the vortex
becomes more confined, but it does not disappear.

{\Large Acknowledgments}

The authors are grateful to M. M. Ferreira Jr. for very clarifying and
detailed discussions. They also express their gratitude to CNPq for the
invaluable financial help.

\end{document}